\documentclass[twocolumn,showpacs,amsmath,amssymb]{revtex4}
\usepackage{graphicx}

\begin{document}

\title
{\Large \bf
 Thermodynamic properties of tunneling quasiparticles  in   graphene-based  structures
}

\author{
Dima Bolmatov$^{a}$\footnote{e-mail: d.bolmatov@qmul.ac.uk}
}
\affiliation{$^{a}$ School of Physics,Queen Mary University of London, Mile End Road, London, E1 4NS, UK 
}


\begin{abstract}
Thermodynamic properties of quasiparticles in a graphene-based  structures are investigated. Two graphene superconducting layers (one superconducting component is placed on the top layered-graphene  structure and the other component in the bottom) separated by oxide dielectric layers and one normal graphene layer in the middle. The quasiparticle flow emerged due to external gate voltage, we considered it as a gas of electron-hole pairs whose components belong to different layers. This is a striking result in view of the complexity of these systems: we have established that specific heat exhibits  universal (-$T^3$) behaviour at low $T$, independent from the gate voltage and the superconducting gap. The experimental observation of this theoretical prediction would be an important step towards our understanding of critical massless matter.

\end{abstract}
\maketitle

\section{Introduction}
Graphene is a unique system in many ways \cite{Novosel-1,Topsakal-1,Sahin-1}. It is truly two-dimensional systems, has unusual electronic excitations described in terms of Dirac fermions that move in a curved space, is an interesting mix of a semiconductor (zero density of states) and a metal (gaplessness), and has properties of soft matter. The electrons in graphene seem to be almost insensitive to disorder and electron-electron interactions and have very long mean free paths \cite{Denis-1}. Hence, graphene's properties are different from what is found in usual metals and semiconductors \cite{GN-1,Sahin-2}. Graphene has also a robust but flexible structure with unusual phonon modes that do not exist in ordinary three-dimensional solids. In some sense, graphene brings together issues in quantum gravity and particle physics, and also from soft and hard condensed matter. Interestingly enough, these properties can be easily modified with the application of electric and magnetic fields, addition of layers, control of its geometry, and chemical doping \cite{Luis-1,Jun-1}. Moreover, graphene can be directly and relatively easily probed by various scanning probe techniques from mesoscopic down to atomic scales, because
it is not buried inside a 3D structure \cite{Pan-1}. This makes graphene one of the most versatile systems in condensed-matter research.
\begin{figure}
	\centering
\includegraphics[scale=0.34]{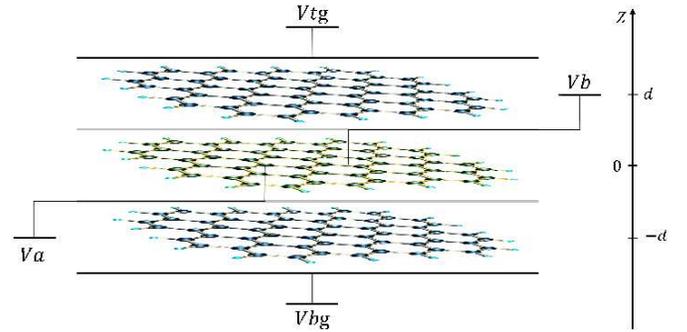}  
	\caption{Graphene multilayer device tunneling structure. Two sheets of superconducting
graphene are separated by thin dielectric oxide layers and one normal graphene monolayer in the middle. Separate gate
electrodes make it possible to vary independently the
carrier concentration in the normal and superconducting
 graphene layers.}
	\label{fig-1}
\end{figure}
Besides the unusual basic properties, graphene has the potential for a large number of applications \cite{Geim-1}, from chemical sensors \cite{Chen-1} to transistors \cite{Nel-1,Oos-1}. Graphene can be chemically and/or structurally modified in order
to change its functionality and henceforth its potential applications. Moreover, graphene can be easily obtained from graphite, a material that is abundant on the Earth's
surface. This particular characteristic makes graphene one of the most readily available materials for basic research. Whereas many papers have been written on monolayer graphene in the past few years, only a small fraction actually deal with multilayers \cite{Haas-1,Yan-1}. The majority of the
theoretical and experimental efforts have concentrated on the single layer, perhaps because of its simplicity and the natural attraction that a one atom thick material,
which can be produced by simple methods in almost any laboratory, creates. Nevertheless, few-layer graphene is equally interesting and unusual with a technological potential,
perhaps larger than the single layer \cite{Bol-1}. Indeed, the theoretical understanding and experimental exploration of multilayers is far behind the single layer. 

Graphene can be considered as a semiconductor with zero band gap. Electron energy spectrum of graphene contains two Dirac points that separate the electron and the hole subband. In a multilayer structure the Fermi levels of the hybrid system layers can be adjusted independently by the gate voltage. The electron-hole symmetry near the Dirac points ensures perfect nesting between the electron and the hole Fermi surfaces. A flow of electron-hole pairs in the graphene-layered structures is equivalent to two oppositely directed electrical currents in the layers. Therefore, the flow of such pairs is a kind of superconductivity \cite{Bol-2}. It is believed that electron-hole pairs may demonstrate superfluid behaviour \cite{Hon-1,Fil-1}. In this letter we consider multilayer graphene structure and claim: flow of electron-hole pairs consisting of an electron from the top (bottom) graphene layer and a hole from the bottom (top) graphene layer behave as superconductive one. This phenomenon can be obtained in a graphene-layered structure with  dielectric layers, pairing of electrons in one top (bottom) graphene layer with holes in the bottom (top) graphene layer occurs.

In the weak-coupling limit, exciton condensation is a consequence of the Cooper    instability \cite{Kel-1} of solids with occupied conduction-band states and empty valence-band states inside identical Fermi surfaces. Each layer has two Dirac-cone bands centered at inequivalent points in its Brillouin zone.  The particle-hole symmetry of the Dirac equation ensures perfect nesting: the nesting condition requires only that the Fermi surfaces be identical in area and shape and not that layers have aligned honeycomb lattices and hence aligned Brillouin zones, between the electron Fermi spheres in one layer and its hole counterparts in the opposite layer, thereby driving the Cooper instability. Global wave vector mismatches can be removed by gauge transformations. When weak inter-valley electron-electron scattering processes are included only simultaneous momentum shifts of both valleys in a layer are allowed.  
\section{Graphene-based  Structures}
Multilayered structures are the building blocks of many of the most advanced devices presently being developed and produced. They are essential elements of the highest-performance optical sources and detectors, and are being employed increasingly in high-speed and high-frequency digital and analog devices. The usefulness of such structures is that they offer precise control over the states and motions of charge carriers. 

\begin{figure}
	\centering
\includegraphics[scale=0.45]{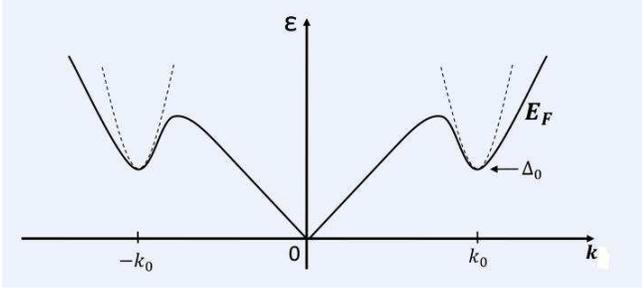}  
	\caption{Scheme of dispersion relation of elementary excitations in graphene multilayer device tunneling structure. The spectrum of elementary excitations of quasiparticles in graphene-based structure was proposed in the He$^{4}$-like manner: the linear portion near $k=0$ represents "phonons" (due to single normal graphene layer in the middle of structure) and the portion near $k=k_{0}$ corresponds to "rotons", which requires a minimal energy $\Delta_{0}$ for its creation.}
	\label{fig-2}
\end{figure}

In Fig.\ref{fig-1} we illustrate a device geometry in which flow  of electron-hole pairs may be observed and in Fig.\ref{fig-2} we sketchily represent the dispersion relation of elementary excitations in graphene multilayer device tunneling structure. We have two superconducting graphene monolayers which are separated by  thin oxide barriers and one normal graphene monolayer in the middle. Top and bottom gates ($V_{tg}$ and $V_{bg}$) are used to electrostatically. The top and bottom gates are separated manipulate the quasiparticle concentrations in the top and bottom layers. The top and bottom gates are separated from the graphene layers by gate oxides which are several
nanometers thick. As the phenomenon of superfluidity of electron-hole pairs  has not yet been experimentally observed in graphene multilayer structures.

A dissipationless flow of electron-hole pairs
in equilibrium through the graphene mono-layer and thin oxide layers, depending on the
phase difference between the two superconducting graphene layers. The single-particle Hamiltonian in graphene is the two-dimensional Dirac Hamiltonian
 \[ H=\left( \begin{array}{c}
\begin{array}{cc}
H_{+} & 0 \\ 0 & H_{-}
 \end{array}
 \end{array}
 \right)
 \]
 where $H_{\pm}=-i\hbar\upsilon_{F}(\sigma_{x}\partial_{x}\pm\sigma_{y}\partial_{y}+U)$  acting on a four-dimensional spinor ($\Psi_{A+}$,$\Psi_{B+}$,$\Psi_{A-}$,$\Psi_{B-}$). The indices $A, B$ label the two sublattices of the honeycomb lattice of carbon atoms, while
the indices $\pm$ label the two valleys of the band structure. There is an additional spin degree of freedom, which plays
no role here. The 2$\times$2 Pauli matrices $\sigma_{i}$ act on the
sublattice index.

The time-reversal operator interchanges the valleys
\[ \Upsilon=\left( \begin{array}{c}
\begin{array}{cc}
0 & \sigma_{z} \\ \sigma_{z} & 0
 \end{array}
 \end{array}
 \right)\wp =\Upsilon^{-1}
 \]
with $\wp$ the operator of complex conjugation. We consider a sheet of graphene in the $x-y$ plane. Electron and hole excitations are described by the Bogoliubov-de Gennes equation
\[ \left( \begin{array}{c}
\begin{array}{cc}
H-E_{F} & \Delta \\ \Delta^{*} & E_{F}-\Upsilon H\Upsilon^{-1}
 \end{array}
 \end{array}
 \right)\left( \begin{array}{c}
\begin{array}{cc}
u \\ v
 \end{array}
 \end{array}
 \right)=\varepsilon\left( \begin{array}{c}
\begin{array}{cc}
u \\ v
 \end{array}
 \end{array}
 \right) \]
with $u$ and $v$ the electron and hole wave functions, $\varepsilon>0$ the excitation energy (relative to the Fermi energy $E_{F}$), $H$ the single-particle Hamiltonian, and $\Upsilon$ the time-reversal operator. The pair potential $\Delta(\bold{r})$ couples time-reversed electron and hole states. In the absence of a magnetic field, the Hamiltonian is time-reversal invariant, $\Upsilon H\Upsilon^{-1}=H$ and we yield two decoupled sets of four equations each, of the form
\[ \left( \begin{array}{c}
\begin{array}{cc}
H_{\pm}-E_{F} & \Delta \\ \Delta^{*} & E_{F}-H_{\pm}
 \end{array}
 \end{array}
 \right)\left( \begin{array}{c}
\begin{array}{cc}
u \\ v
 \end{array}
 \end{array}
 \right)=\varepsilon\left( \begin{array}{c}
\begin{array}{cc}
u \\ v
 \end{array}
 \end{array}
 \right) \]

Separate gate electrodes make it possible to vary independently the carrier concentration of electron-hole pairs in the normal direction ($z$-direction) to graphene mono-layer, thin oxide layers and superconducting graphene layers. For $-d<z<d$, the pair potential vanishes identically, disregarding
any intrinsic superconductivity of graphene, where $d$ is a total width of graphene mono-layer and thin oxide layers. For $z<-d$ and $z>d$ the superconducting graphene layers will induce a nonzero pair potential $\Delta(z)$ via the proximity effect similarly to what happens in a planar junction between a two-dimensional electron gas and a superconductor\cite{Volkov-1}. The bulk value $\Delta_{0}e^{i\phi}$ for $z<-d$ and $\Delta_{0}e^{-i\phi}$ for $z>d$ (with $\phi$ the superconducting phase) is reached at a distance from the
interface which becomes negligibly small if the Fermi wavelength $\lambda_{F}^{'}$ in superconducting layers is much smaller than the value $\lambda_{F}$ in graphene mono-layer and thin oxide layers. We assume that the electrostatic potential $U$ in graphene mono-layer, thin oxide and superconducting graphene layers may be adjusted independently by a gate voltage or by doping. Since the zero of potential is arbitrary, we may take: $U(\bold{r})=0$ for $-d<z<d$ and $U(\bold{r})=-U_{0}$ otherwise. For $U_{0}$ large positive, and $E_{F}\geq 0$, the Fermi wave vector $k_{F}^{'}\equiv 2\pi/\lambda_{F}^{'}=(E_{F}+U_{0})/\hbar\upsilon_{F}$ in graphene superconducting layers is large compared to the value $k_{F}\equiv 2\pi/\lambda_{F}=E_{F}/\hbar\upsilon_{F}$ in graphene mono-layer and thin oxide layers (with $\upsilon_{F}$ the energy-independent velocity in graphene).

\section{Specific heat}
For the graphene mono-layer and oxide thin layers (where $\Delta=0=U$) we assume that Fermi level is tuned to the point of zero carrier concentration. In the superconducting graphene layers there is a gap in the spectrum of magnitude $|\Delta|=\Delta_{0}$.

The canonical partition function for this graphene-based structure can be interpreted as a collection of normal mode oscillators, the oscillator labelled by $\bold{k}$ containing $n_{\bold{k}}$ quanta of energy $\varepsilon(\bold{k})$. First we calculate the Helmholtz free energy for normal graphene mono-layer ($phonon$-like dependence) and thin oxide layers (its make dispersion relation between linear and quadratic dependence smooth, see Fig.\ref{fig-2}) which can be written as
\begin{eqnarray}
F_{n}(T,V)=-2Nk_{B}TV\int_{-k_{d}}^{k_{d}}\frac{dk}{2\pi} k\ln{\left[1+\exp{\left(-\frac{\hbar\upsilon_{F}k}{k_{B}T}\right)}\right]}
\end{eqnarray}
where $k_{d}$ ($-k_{d}$) is the largest (lowest) wavevector for which the linear dispersion is a reasonable approximation. Therefore, the contribution to free energy from the $normal$ region of graphene-based structure is
\begin{eqnarray}
F_{n}(T,V)\simeq -\frac{3\zeta(3)Nk_{B}T^{3}V}{2\pi\hbar^{2}\upsilon_{F}^{2}}
\end{eqnarray}
where $N$ is the number of the $2$ component Dirac flavors, $N = 4$ in the single layer graphene and $\zeta(n)$ is the Riemann zeta function. Contribution to the specific heat per unit volume is
\begin{eqnarray}
C_{V}^{n}=-\frac{T}{V}\left( \frac{\partial^{2}F_{n}}{\partial T^2}\right)_{V}=\frac{43.2k_{B}T^{2}}{\pi\hbar^{2}\upsilon_{F}^{2}}
\end{eqnarray} 
\subsection{Specific heat of superconducting layers: graphene-based layers}
The grand potential of the system ($roton$-like dependence) is given by
\begin{eqnarray}
\nonumber
q(V,T)\equiv -\beta F_{s}(V,T)=-\sum_{\bold{k}}\ln{[1-\exp{(-\beta\varepsilon(\bold{k}))}]}\\
\simeq\sum_{\bold{k}}\exp{(-\beta\varepsilon(\bold{k}))}\simeq \bar{N}
\end{eqnarray}
where $\bar{N}$ is the "equilibrium" number of $rotons$ in the superconducting graphene layers. The summation over $\bold{k}$ may be replaced by an integration, with the result
\begin{eqnarray}
\nonumber
F_{s} &=& -4k_{B}T\bar{N}\\
&=& -8k_{B}TV\int_{k_{1}}^{k_{2}}\frac{dk}{2\pi}ke^{-\beta(\Delta+U+\hbar\upsilon_{F}(k-k_{0}))}
\end{eqnarray}
where $k_{1}$ and $k_{2}$ are the lowest and highest values respectively of $k$ for which the quadratic approximation to the dispersion curve is reasonable (the limits of integration may be extended to $\pm\infty$). The free energy of the $roton$-like superfluid gas is given by
\begin{eqnarray}
F_{s}\simeq -\frac{4(k_{B}T)^{2}V}{\pi\hbar\upsilon_{F}}\left(\frac{k_{B}T}{\hbar\upsilon_{F}}-k_{0}\right)e^{-(2\Delta_{0}\cos{\phi}+U)/k_{B}T}
\end{eqnarray}
The $roton$-like contribution to the specific heat we estimate as: $C_{V}^{s}=C_{V,1}^{s}+C_{V,2}^{s}$, where
\begin{eqnarray}
\nonumber
C_{V,1}^{s}=\frac{4k_{B}(k_{B}T)^2}{\pi(\hbar\upsilon_{F})^2} e^{-(2\Delta_{0}\cos{\phi}+U)/k_{B}T}\times\\
\left[6+4\cdot\frac{2\Delta_{0}\cos{\phi}+U}{k_{B}T}+\left(\frac{2\Delta_{0}\cos{\phi}+U}{k_{B}T}\right)^{2}\right]\\
\nonumber
C_{V,2}^{s}=-\frac{4k_{B}k_{0}(k_{B}T)}{\pi\hbar\upsilon_{F}} e^{-(2\Delta_{0}\cos{\phi}+U)/k_{B}T}\times\\
\left[2+2\cdot\frac{2\Delta_{0}\cos{\phi}+U}{k_{B}T}+\left(\frac{2\Delta_{0}\cos{\phi}+U}{k_{B}T}\right)^{2}\right]
\end{eqnarray}
\begin{figure}
	\centering
\includegraphics[scale=0.7]{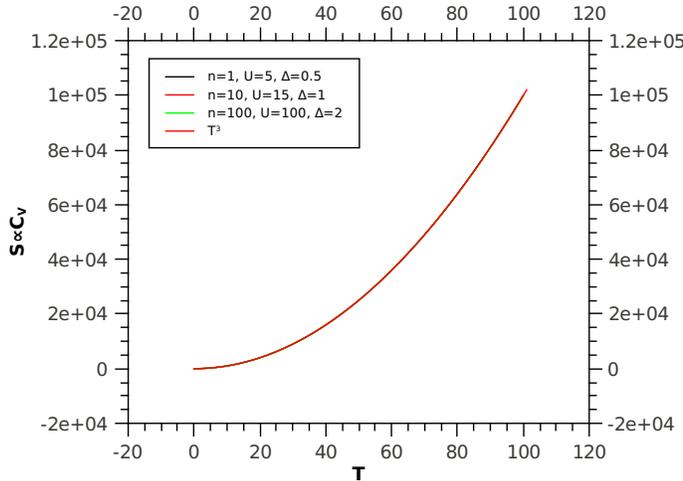}  
	\caption{Dependence of specific heat versus temperature exhibits Debye-like behaviour ($T^3$-like). This is a striking result in view of all four curves sit on top of each other.}
	\label{fig-3}
\end{figure}
where $k_{0}=n/d$, $d=0.4$ nm is the thickness of graphene sheet, n is the number of oxide dielectric layers (the thickness of each of them is presumed equivalent to graphene one) between normal and superconducting graphene layers. Rough low-temperature estimates often quote the result
\begin{eqnarray}
S\propto C_{V}^{n}+C_{V}^{s}
\end{eqnarray}
where $S$ is thermopower \cite{Scar-1}. In Fig.\ref{fig-3} is depicted the specific heat against temperature. It is remarkable that in graphene-based superconducting structure the phonon-assisted drag effect is suppressed. The carrier concentration can be tuned varying different parameters: the thickness of oxide layers, phase difference in superconducting layers, pair and electrostatic potentials on one hand and on the another proposed structure exhibits Debye-like behaviour.
\section{Conclusion}
The practical significance of this investigation rests on
the expectation that high-quality contacts between a superconducting
graphene and normal graphene sheets can be realized. This expectation is supported by the experience with carbon nanotubes (rolled up sheets of graphene), which have
been contacted succesfully by superconducting electrodes. Graphene-based structures provides a unique opportunity to explore the physics of the
"relativistic Josephson effect", which had remained unexplored
in earlier work on relativistic effects in high-temperature
and heavy-fermion superconductors.

We believe that this device structure can also lead to greater understanding in another condensed matter systems such as high-temperature superconductors and exciton gases. The low cost experimental techniques allow for easy verification of the proposed hypothesis.
\section{ACKNOWLEDGMENT}
Authors are very indebted to C. Y. Mou, E. Babaev, M. M. Parish, M. Baxendale and  D. J. Dunstan for stimulating discussions and fruitful suggestions. We acknowledge Myerscough Bequest and School of Physics at Queen Mary University for financial support.

\end{document}